**High-pass filtered fidelity-imposed network edit (HP-FINE) for robust quantitative susceptibility mapping from high-pass filtered phase**

Jinwei Zhang[1,2], Alexey Dimov[2], Chao Li[2,4], Hang Zhang[3], Thanh D. Nguyen[2], Pascal Spincemaille[2], Yi Wang[1,2]

[1]Meinig School of Biomedical Engineering, Cornell University, Ithaca, NY

[2]Department of Radiology, Weill Cornell Medicine, New York, NY

[3]Department of Electrical and Computer Engineering, Cornell University, Ithaca, NY

[4]Department of Applied Physics, Cornell University, Ithaca, NY

Correspondence:

Yi Wang, PhD.

Radiology, Weill Cornell Medicine

407 E 61st St, New York, NY 10065, USA

E-mail: yiwang@med.cornell.edu

## ABSTRACT

**Purpose:** To improve the generalization ability of deep learning based predictions of quantitative susceptibility mapping (QSM) from high-pass filtered phase (HPFP) data.

**Methods:** A network fine-tuning step called HP-FINE is proposed, which is based on the high-pass filtering forward model with low-frequency preservation regularization. Several comparisons were conducted: 1. HP-FINE with and without low-frequency regularization, 2. three 3D network architectures (Unet, Progressive Unet, and Big Unet), 3. two types of network output (recovered field and susceptibility), and 4. pre-training with and without the filtering augmentation. HPFP datasets with diverse high-pass filters, another acquisition voxel size, and prospective acquisition were used to assess the accuracy of QSM predictions. In the retrospective datasets, quantitative metrics (PSNR, SSIM, RMSE and HFEN) were used for evaluation. In the prospective dataset, statistics of ROI linear regression and Bland-Altman analysis were used for evaluation.

**Results:** In the retrospective datasets, adding low-frequency regularization in HP-FINE substantially improved prediction accuracy compared to the pre-trained results, especially when combined with the filtering augmentation and recovered field output. In the prospective datasets, HP-FINE with low-frequency regularization and recovered field output demonstrated the preservation of ROI values, a result that was not achieved when using susceptibility as the output. Furthermore, Progressive Unet pre-trained with a combination of multiple losses outperformed both Unet and Progressive Unet pre-trained with a single loss in terms of preserving ROI values.

**Conclusion:** Experiments in both retrospective and prospective in vivo data demonstrate improved accuracy in QSM reconstruction using HP-FINE. Our code is available at https://github.com/Jinwei1209/SWI_to_QSM/

## INTRODUCTION

Susceptibility-weighted imaging (SWI) (1) is a widely used MRI method to visualize susceptibility sources, such as veins, calcifications and hemorrhages, by showing the effect of their tissue-induced magnetic field. SWI is obtained by combining the high-pass filtered phase (HPFP) and magnitude images of the gradient echo (GRE) data. The HPFP image is computed as the angle of the complex division of the original GRE by its low-pass filtered version. This operation suppresses the much larger background field which predominantly consists of low spatial frequency components and impedes clear visualization of the local tissue field (2). SWI has been used in clinical applications such as stroke, cerebrovascular disease and neurodegenerative disorders (3). However, this technique does not allow a direct quantitative measurement of the underlying tissue susceptibility.

Quantitative susceptibility mapping (QSM) (4) has been proposed to map tissue susceptibility by solving a dipole inversion problem. Typically, the tissue-induced field map is first obtained by fitting the phase of the complex multi-echo GRE data (5,6) followed by background field removal (7-10). Then, a regularized dipole inversion is performed to estimate the underlying tissue susceptibility distribution (11-14).

To date, a large amount of SWI HPFP data has been collected for which the underlying complex GRE data is no longer available. Despite potential interest in their quantitative analysis, accumulated HPFP images are incompatible with QSM reconstruction due to the removal of the low spatial frequency components of the tissue field, which is known to be a suboptimal method for the removal of background field as it removes some of the tissue field as well. Thanks to the rapid development of convolutional neural networks (CNNs) in medical imaging, there is an increasing interest in recovering QSM from HPFP data in SWI using CNNs (15-18). In one approach, a Unet architecture (19,20) was applied to recover unfiltered phase data from HPFP data and compute QSM using the recovered phase data (16). A second approach reconstructed QSM directly from HPFP data, either using a network architecture with a dilated interaction block (18), an edge prior and a cross dense block (15) or a generative adversarial network (GAN) (17). These promising results were obtained with the assumption that training and test data were acquired with the same imaging parameters (17,18). However, generalization errors might occur when tested with data obtained by a different acquisition protocol, such as a different voxel size (15) or an unseen/unknown high-pass filter (16). To improve generalizability of unseen filters, (16) proposed to augment the training data with a diverse set of high-pass filters and cutoffs. The study showed promising results when tested with filters unseen during training.

In this paper, we introduce a novel test-time fine-tuning strategy aimed at reducing the generalization errors caused by both varying voxel sizes and unknown high-pass filters. We incorporate the forward high-pass filtering operations into the loss function with a regularization term preserving the low-frequency information of the network output for subject-specific test-time fine-tuning (21,22). Furthermore, we implement and compare several network architectures including Unet, Progressive Unet (concatenation of multiple Unets) and Big Unet. We also employ and compare recovered field and susceptibility as two types of network output.

## METHODS

### Data acquisition and preprocessing

Data were acquired following an IRB approved protocol. All images used in this work were de-identified to protect the privacy of human participants.

Retrospective dataset one

Multi-echo 3D GRE data acquired from 30 multiple sclerosis (MS) patients on a clinical 3T scanner (Siemens Healthineers) were retrospectively analyzed. The acquisition parameters were as follows: FA=15°, FOV = 24.0 cm, $TE_1$ = 6.3 ms, TR = 50.0 ms, #TE = 10, $\Delta$TE = 4.1 ms, acquisition matrix = 320 × 320 × 48, voxel size = 0.75 × 0.75 × 3 $mm^3$. The tissue field was estimated using non-linear fitting of the multi-echo phase data (6), followed by phase unwrapping and background field removal (7). QSM was reconstructed by Morphology Enabled Dipole Inversion (MEDI) (23) and used as the ground truth. The HPFP $f_{HPF}$ at $TE_4$ = 22.7 ms was computed retrospectively as the Homodyne filtering of the original complex image of slice $c$: $f_{HPF} = arg\left(\frac{c}{L_{FC}(c)}\right)$, where $arg$ is the complex argument and $L_{FC}(\cdot)$ is the low-pass filter with a cutoff frequency $FC$ in Fourier space. Three types of low-pass filters are presented in Appendix A. Data from 18/2/10 subjects were used as training/validation/test datasets.

Retrospective dataset two

Additionally, multi-echo 3D GRE data were acquired from 10 healthy volunteers on another scanner (GE Healthcare) as a second test dataset with a different acquisition voxel size. The acquisition parameters were as follows: FA=15°, FOV = 25.6 cm, $TE_1$ = 2.0 ms, TR = 36.0 ms, #TE = 10, $\Delta$TE = 3.4 ms, acquisition matrix = 256 × 206 × 80, voxel size = 1 × 1 × 2 $mm^3$. QSM and HPFP $f_{HPF}$ at $TE_6$ = 22.4 ms were computed similarly to the first dataset.

Prospective dataset

3D GRE data was acquired prospectively from 9 healthy volunteers with both single-echo (TE = 22.7 ms, TR = 50.0 ms) and multi-echo acquisitions using the same 3T Siemens scanner and imaging parameters identical to the retrospective HPFP dataset. HPFP and SWI images from the single-echo GRE data were generated prospectively by the scanner software with unknown high-pass filter parameters. QSM was reconstructed from the multi-echo GRE data using the same pipeline as the retrospective HPFP dataset. This formed a third test set with prospective HPFP inputs and their corresponding QSM references.

**Network architectures, pre-training and fine-tuning**

Network architectures

HPFP was used as the network input. Recovered field (relative difference field, RDF) and susceptibility (QSM) were used as two types of network outputs and their performances were compared. MEDI (23) was used to produce QSM from the RDF output. Three 3D network architectures were used and compared: Unet, Progressive Unet (Prognet), and Big Unet (with the same number of parameters as Prognet). Prognet offers advantages such as increased capacity compared to Unet and the feasibility of memory-constrained fine-tuning via updating only the last Unet of Prognet, a process not straightforwardly applicable in Big Unet. A detained description of the three network architectures is presented in Appendix B.

Pre-training

3D patches with patch size 128*128*32 were randomly extracted during pre-training. $L_1$ loss between the ground truth and predicted RDFs or QSMs was used for supervised pretraining of all networks with a batch size of 4. In Prognet, a sum of $L_1$ losses from each intermediate output were also applied and compared. Additionally, 4-Unet progression in Prognet was used to maximize GPU memory usage during pre-training (11GB, RTX 2080Ti). Adam optimizer (24) was deployed with $10^{-4}$ learning rate and 20000 iterations.

Data augmentation using three types of filters with $FCs = 1/4, 3/8, 1/2, 5/8,$ and $3/4$ in Appendix A was optionally applied to all networks for pre-training. For comparison, pre-training without such augmentation solely used the Hann filter with $FC = 3/8$.

High-pass filtered fidelity-imposed network edit (HP-FINE)

Whole 3D volumes were fed into networks for testing. For the QSM output with an initial prediction $\chi_0$ from the pre-trained networks, the following high-pass filtering dipole convolution forward model with low-frequency preservation regularization was applied to fine-tune the pre-trained network weights during inference:

$$loss_{QSM}(\chi) = \left\| arg\left(\frac{m\cdot e^{i(d*\chi)}}{L_{FC}(m\cdot e^{i(d*\chi)})}\right) - f_{HPF}\right\|_2^2 + \lambda\left\|L_{FC}\left(m\cdot e^{i(d*\chi)}\right) - L_{FC}\left(m\cdot e^{i(d*\chi_0)}\right)\right\|_2^2, \quad (1)$$

where $\chi$ is the QSM output, $d$ is the dipole kernel, $m$ is the magnitude image at the corresponding SWI echo time, and $f_{HPF}$ is the HPFP input. For the RDF output with an initial prediction $f_0$ from the pre-trained networks, similar loss was applied replacing the dipole convolution with the unfiltered field:

$$loss_{RDF}(f) = \left\| arg\left(\frac{m\cdot e^{if}}{L_{FC}(m\cdot e^{if})}\right) - f_{HPF}\right\|_2^2 + \lambda\left\|L_{FC}\left(m\cdot e^{if}\right) - L_{FC}\left(m\cdot e^{if_0}\right)\right\|_2^2. \quad (2)$$

$L_{FC}(\cdot)$ was chosen as the Hann filter with $FC = 3/8$ in Eqs. 1 and 2, considering a priori unknown filter type and cutoff frequency in practical situations. HP-FINE backpropagation using Eqs. 1 or 2 and Adam optimizer with $10^{-4}$ learning rate was terminated when the relative change of the loss between two consecutive iterations fell below $5\times10^{-3}$ (22) or started to fluctuate. For Big Unet, HP-FINE was not applied due to memory constraint for whole 3D volume inference. For Prognet, HP-FINE was applied only to the last Unet.

**Experiments**

The following test datasets were generated for performance evaluation:

1. HPFP test data with various filters were computed from the retrospective test dataset one using the Hann and Gaussian filters with $FCs = 1/4, 3/8, 1/2, 5/8,$ and $3/4$.

2. HPFP test data acquired with another voxel size were also computed from the retrospective test dataset two using the Hann filter with $FC = 3/8$.

3. HPFP test data were also computed by the scanner software from the prospective test dataset.

For the retrospective HPFP test datasets, voxel-wise metrics root mean squared error (RMSE), peak signal-to-noise ratio (PSNR), structure similarity index (SSIM) and high-frequency error norm (HFEN) metrics were used to quantify the QSM reconstruction accuracy (25). For the prospective HPFP test dataset, regions of interest (ROIs) in the left and right caudate nuclei (CN), globus pallidus (GP) and putamen (PU) were manually segmented. For each ROI, the slope and correlation coefficient of linear regression, as well as the bias and 95% limits of agreement of Bland-Altman analysis between the predicted and reference QSM values were reported.

**RESULTS**

An optimal $\lambda = 10^{-4}$ was selected via a grid search to optimize the QSM prediction accuracy on the retrospective validation dataset using the Hann filter with $FC = 3/4$, and then fixed for all subsequent test experiments.

Supplementary Table S1 shows the number of parameters and computational time for various architectures. An increase in computational time averaging 12 seconds was observed on Unet and Prognet during HP-FINE using Eqs. 1 or 2, compared to an average of 1 second for inference without fine-tuning.

Figure 1 shows the predicted QSMs generated from the RDF outputs of pre-trained (1a), HP-FINE with $\lambda = 0$ (1b), and HP-FINE with $\lambda = 10^{-4}$ (1c) on the test data using the Hann filter with $FC = 3/8$. The filtering augmentation was applied to all networks during their pre-training. The errors of Prognet, Unet, Prognet (single loss), and Big Unet were comparable in the pre-trained results, while the errors of Prognet and Unet were reduced in the HP-FINE with $\lambda = 10^{-4}$ results. HP-FINE with $\lambda = 0$ on Prognet, Unet, and Prognet (single loss) and $\lambda = 10^{-4}$ on Prognet (single loss) exhibited moderate underestimation of GP.

Supplementary Figures S1 and S2 add comparisons between the RDF and QSM outputs, as well as comparisons between pre-training with and without the filtering augmentation, using the Hann filters with $FC = 1/4$ and $3/4$. For the RDF output, both the filtering augmentation and HP-FINE with $\lambda = 10^{-4}$ contributed to improving accuracy. For the QSM output, the filtering augmentation improved accuracy, whereas HP-FINE resulted in moderate underestimation of GP.

Supplementary Figure S5 shows the boxplot of quantitative metrics using the Hann filters with $FCs = 1/4, 3/8, 1/2, 5/8,$ and $3/4$. Overall, the filtering augmentation, RDF output, HP-FINE with $\lambda = 0$, and HP-FINE with $\lambda = 10^{-4}$ all contributed to improved accuracy. Among these, HP-FINE made a more substantial contribution by improving from the pre-trained results, with HP-FINE at $\lambda = 10^{-4}$ slightly outperforming $\lambda = 0$ in most cases for the RDF output.

Figure 2 and supplementary Figures S3, S4 and S6 show reconstructions and quantitative metrics on the test data using the Gaussian filters with $FCs = 1/4, 3/8, 1/2, 5/8,$ and $3/4$. Similar results to those of the Hann filters were observed.

Figure 3 shows the predicted QSMs of pre-trained (3a), HP-FINE with $\lambda = 0$ (3b), and HP-FINE with $\lambda = 10^{-4}$ (3c) on the test data acquired with another voxel size. The filtering augmentation was applied to all networks during their pre-training. Visual blurriness was observed in the pre-trained results but was reduced in both HP-FINE with $\lambda = 0$ and $\lambda = 10^{-4}$. The RDF output showed lower errors compared to the QSM output for each network and reconstruction method.

Supplementary Figure S7 shows the boxplot of quantitative metrics on the test data acquired with another voxel size. Compared to the RDF output, all methods exhibited a noticeable performance decrease with the QSM output. HP-FINE with $\lambda = 0$ and $\lambda = 10^{-4}$ significantly improved the accuracy of all pre-trained networks. Best performing methods were the ones using the RDF output and HP-FINE with $\lambda = 10^{-4}$.

Figure 4 shows the predicted QSMs of pre-trained (4a), HP-FINE with $\lambda = 0$ (4b), and HP-FINE with $\lambda = 10^{-4}$ (4c) on the test data generated prospectively by the scanner software. The filtering augmentation was applied to all networks during their pre-training. Moderate blurriness in the pre-trained results was reduced in both HP-FINE with $\lambda = 0$ and $\lambda = 10^{-4}$. The RDF output with HP-FINE at $\lambda = 10^{-4}$ also preserved the deep grey matter susceptibility values after fine-tuning.

Figure 5 shows the statistics of ROI linear regression and Bland-Altman analysis conducted on the prospective test dataset using Prognet with the RDF output and filtering augmentation. HP-FINE at $\lambda = 0$ suffered from under-estimating ROI values, which were preserved by HP-FINE with $\lambda = 10^{-4}$.

Supplementary spreadsheet reports the statistics of all methods on the prospective test dataset. For each network, the RDF output with HP-FINE at $\lambda = 10^{-4}$ outperformed that at $\lambda = 0$ in terms of preserving ROI values, a result not achieved by the QSM output at either $\lambda = 10^{-4}$ or $\lambda = 0$. Prognet outperformed Unet and Prognet (single loss) in the preservation of ROI values.

## DISCUSSION AND CONCLUSION

In this study, we propose a method called HP-FINE to improve the generalization ability of reconstructing QSM from HPFP data using deep learning. Experiments in both retrospective and prospective in vivo data demonstrate improved accuracy in QSM reconstruction after HP-FINE to handle domain shift.

In HP-FINE, $f_{HPF}$ is used in both the network input and fine-tuning losses (Eqs. 1 and 2) in a self-supervised manner. Such fidelity imposed network edit (FINE) approach was proposed in prior work (22) for ill-posed dipole inversion using deep learning (26-31), and has been applied to related inverse problems such as water-fat separation (32). Compared to dipole inversion, QSM reconstruction from HPFP data is a more ill-posed inverse problem, due to properties of both the dipole kernel and the inherent incompleteness of input data. The increased complexity of learning the dipole inversion alongside inverse filtering led to decreased performance compared to learning only the inverse filtering. This was confirmed by the superior performance of the RDF output over the QSM output, as shown in both the pre-trained and HP-FINE results (e.g., Figures 3-4 and supplementary Figures S1-4).

The filtering augmentation for pre-training was proposed in prior work to improve generalization across various filters (16). HP-FINE further reduced generalization errors on top of the filtering augmentation when tested on various unknown filters (supplementary Figures S1-6). HP-FINE also reduced generalization errors caused by a different acquisition voxel size, a challenge that cannot be solely addressed by the filtering augmentation (Figure 3 and supplementary Figures S7). Nevertheless, a more stable HP-FINE step was observed when pre-trained with the filtering augmentation (supplementary Figures S1-7), suggesting that initialized with a more generalizable pre-trained model results in better HP-FINE.

In HP-FINE, when tested on the HPFP data under small to moderate deviations of the filter types and parameters from the ones used in the fine-tune losses (Eqs. 1 and 2), improvement in accuracy was still observed (e.g., the Gaussian filter in Figure 2). A possible explanation is that high-pass filter types and parameters typically selected in practical setting impose very aggressive filtering, preserving only sparse details such as phase edges. Therefore, the exact detail of the low frequency suppression may be less important, leading to a low inconsistency between measured HPFP data and data predicted by a possibly inaccurate forward model. Nevertheless, in instances where more information was filtered out in the HPFP test data (e.g., $FC = 3/4$ in supplementary Figures S2 and S5e), an underestimation of QSM was observed in some HP-FINE results. This could potentially be attributed to the reduced information available or a greater deviation between the forward model in the fine-tune losses and the actual model used for generating the test data.

The low-frequency preservation regularization contributed to improved global quantitative accuracy (supplementary Figures S5-7), meanwhile ensuring ROI value preservation (Figure 5 and the supplementary spreadsheet). Since the HPFP data in the fine-tuning losses only contains high-frequency information, adding low-frequency regularization term penalizes the alteration of the low-frequency information predicted by the pre-trained results. This low-frequency information includes a general depiction of ROIs, which determines the values within these ROIs. However, such low-frequency regularization did not benefit HP-FINE with the QSM output (Figures 3 and 4), potentially because of the increased complexity in predicting QSM rather than recovering the full phase from the HPFP input.

The four architectures used in this study, including two Prognets pre-trained with either a single loss or a sum of multiple losses, exhibited comparable quantitative metrics (supplementary

Figures S5-7). However, better ROI preservation performance was achieved in Prognet pre-trained with multi-loss compared to Unet and Prognet pre-trained with single loss (Figures 1, 2, 4 and supplementary spreadsheet), possibly because of the larger capacity of Prognet over Unet and the more robust intermediate output prediction by imposing multi-loss over single-loss pre-training.

There are several limitations present in this work. First, the efficiency of fine-tuning requires further study. Several approaches can be applied to reduce the fine-tuning time, such as subnet fine-tuning (21) or meta-learning (33). Second, unrolled architectures need to be investigated as another strategy incorporating physical models into deep learning. Unrolled network architectures have become popular in QSM-related research, with applications in multi-echo gradient echo sequence acceleration (31,34,35) and dipole inversion (36-38). Future work should also include implementing the unrolled architecture for phase recovery from HPFP.

## FIGURES

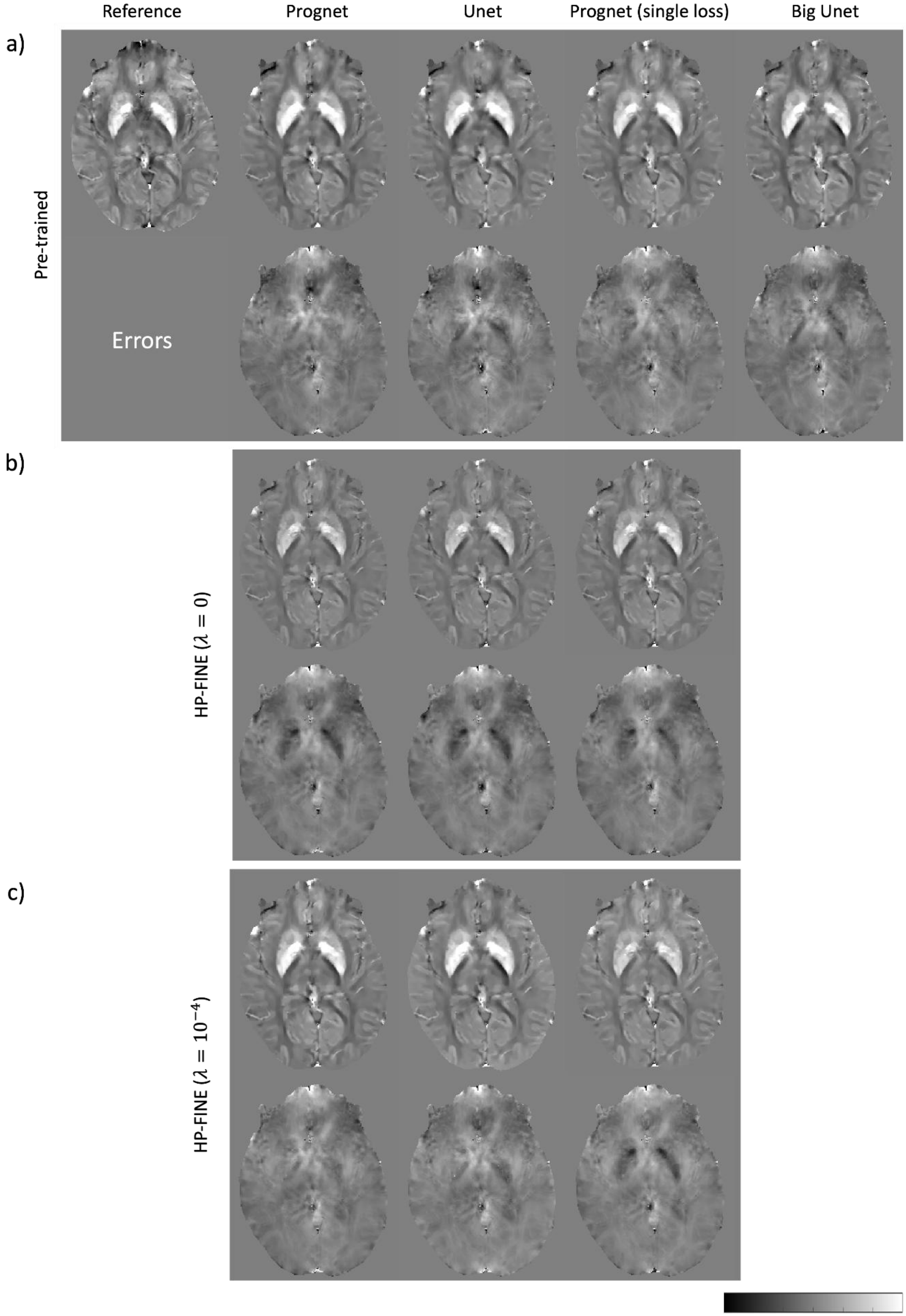

**Figure 1.** QSM predictions generated from the RDF outputs of the test HPFP data using the Hann filter with $FC = 3/8$. Figure a) shows the pre-trained results, while b) shows HP-FINE with $\lambda = 0$, and (c) shows HP-FINE with $\lambda = 10^{-4}$. The filtering augmentation was applied to all networks during their pre-training. Prognet, Unet, Prognet (single loss), and Big Unet exhibited similar errors in the pre-trained results. However, the errors of Prognet and Unet decreased in the HP-FINE with $\lambda = 10^{-4}$ results. Prognet, Unet, and Prognet (single loss) with HP-FINE at $\lambda = 0$ and Prognet (single loss) with HP-FINE at $\lambda = 10^{-4}$ exhibited moderate underestimation of the globus pallidus.

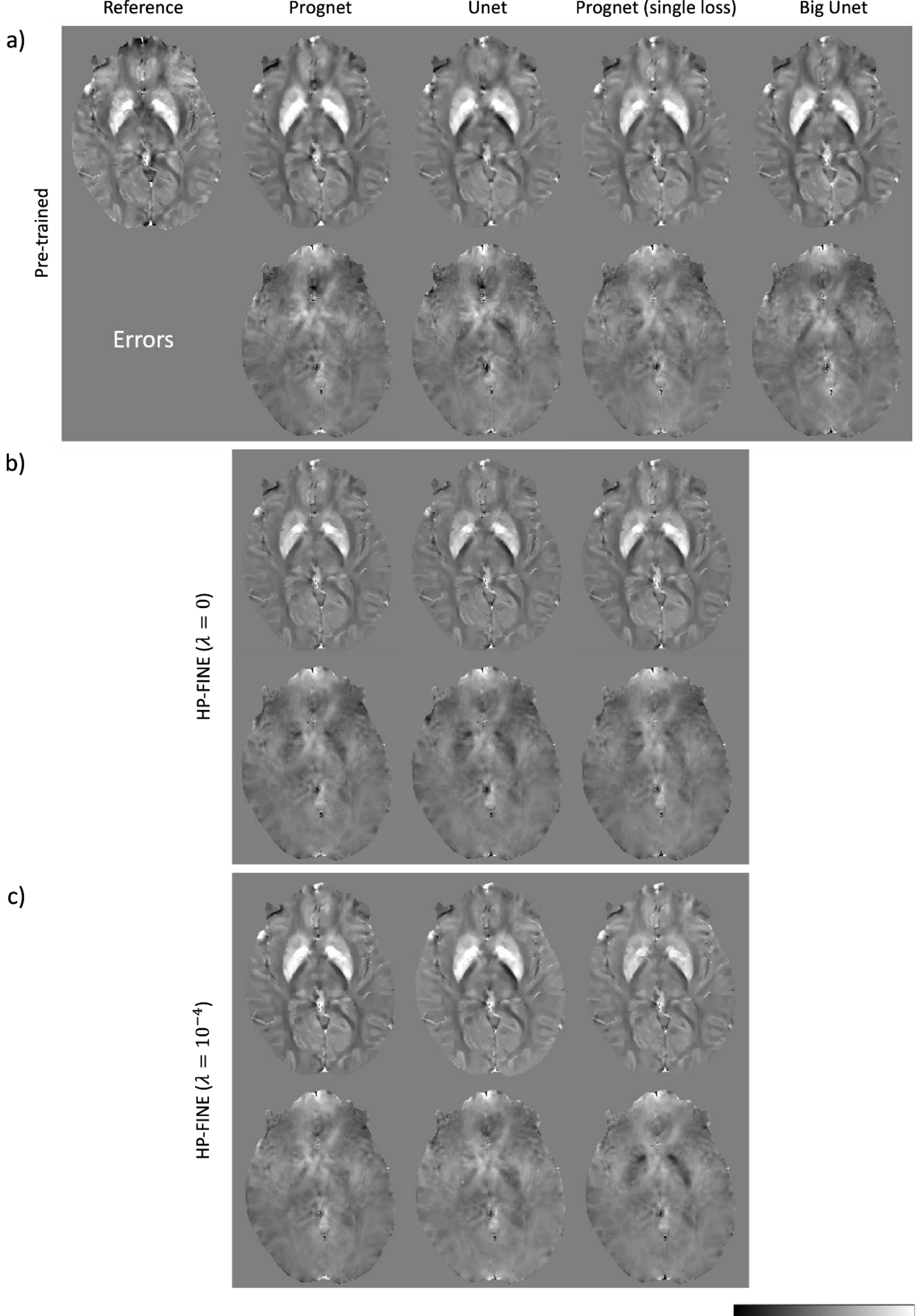

**Figure 2.** QSM predictions of a) the pre-trained, b) HP-FINE with $\lambda = 0$, and c) HP-FINE with $\lambda = 10^{-4}$ results generated from the RDF outputs of the test HPFP data using the Gaussian filter with $FC = 3/8$. The filtering augmentation was applied to all networks during their pre-training. Prognet and Unet exhibited similar errors to Prognet (single loss) and Big Unet in the pre-trained results, but displayed decreased errors in the HP-FINE with $\lambda = 10^{-4}$ results. Small to moderate underestimation of the globus pallidus was observed in Prognet, Unet, and Prognet (single loss) with HP-FINE at $\lambda = 0$ and Prognet (single loss) with HP-FINE at $\lambda = 10^{-4}$.

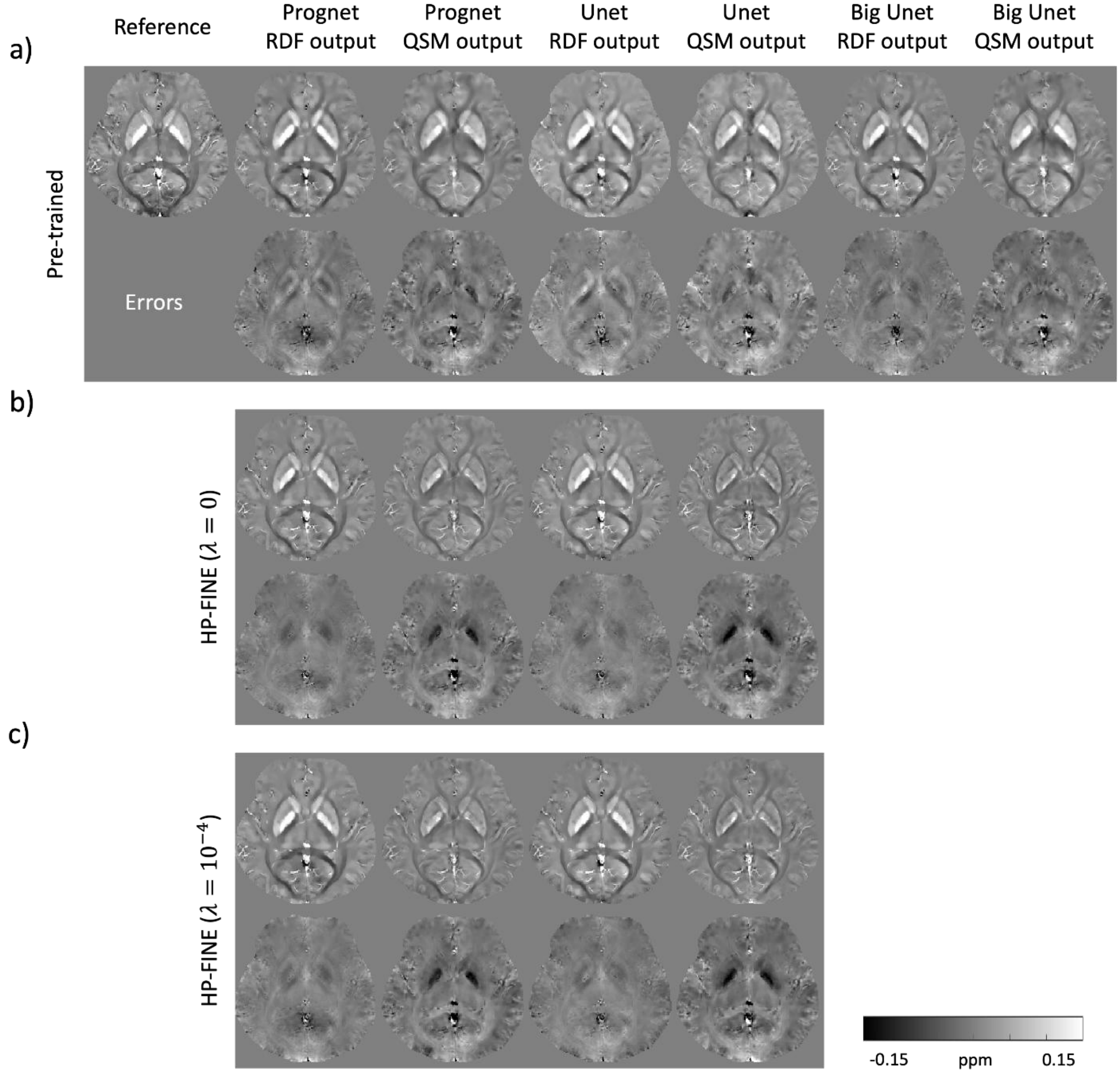

**Figure 3.** QSM predictions of a) the pre-trained, b) HP-FINE with $\lambda = 0$, and c) HP-FINE with $\lambda = 10^{-4}$ results on the test data acquired with another voxel size. The filtering augmentation was applied to all networks during their pre-training. Visual blurriness observed in the pre-trained results was reduced in both HP-FINE with $\lambda = 0$ and $\lambda = 10^{-4}$. For each network and reconstruction method, the RDF output displayed lower errors compared to the QSM output.

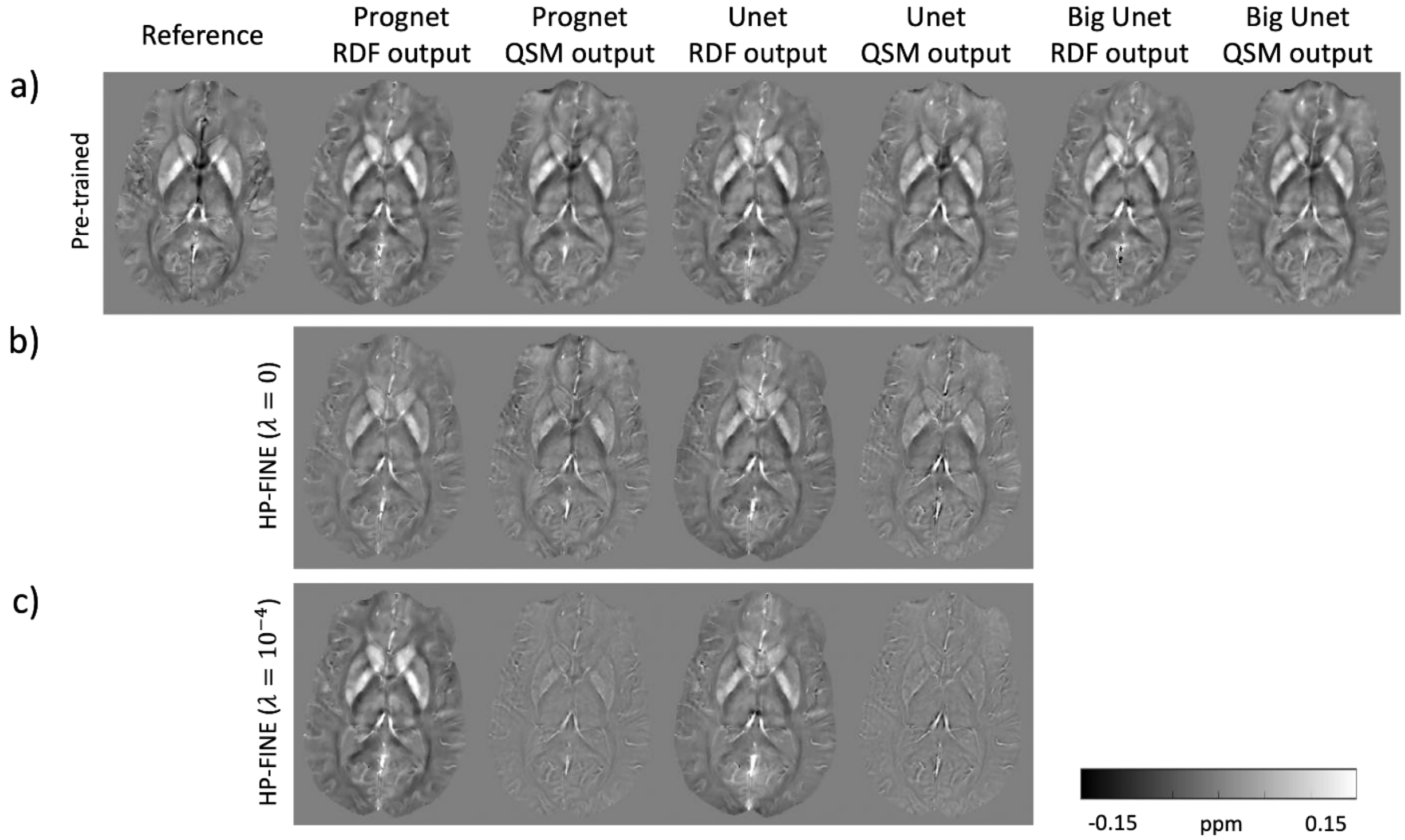

**Figure 4.** QSM predictions of a) the pre-trained, b) HP-FINE with $\lambda = 0$, and c) HP-FINE with $\lambda = 10^{-4}$ on the test data generated prospectively by the scanner software. The filtering augmentation was applied to all networks during their pre-training. Moderate blurriness in the pre-trained results was reduced in both HP-FINE with $\lambda = 0$ and $\lambda = 10^{-4}$. After fine-tuning, both Prognet and Unet with the RDF output and HP-FINE at $\lambda = 10^{-4}$ preserved the deep grey matter susceptibility values, with Prognet performing better than Unet.

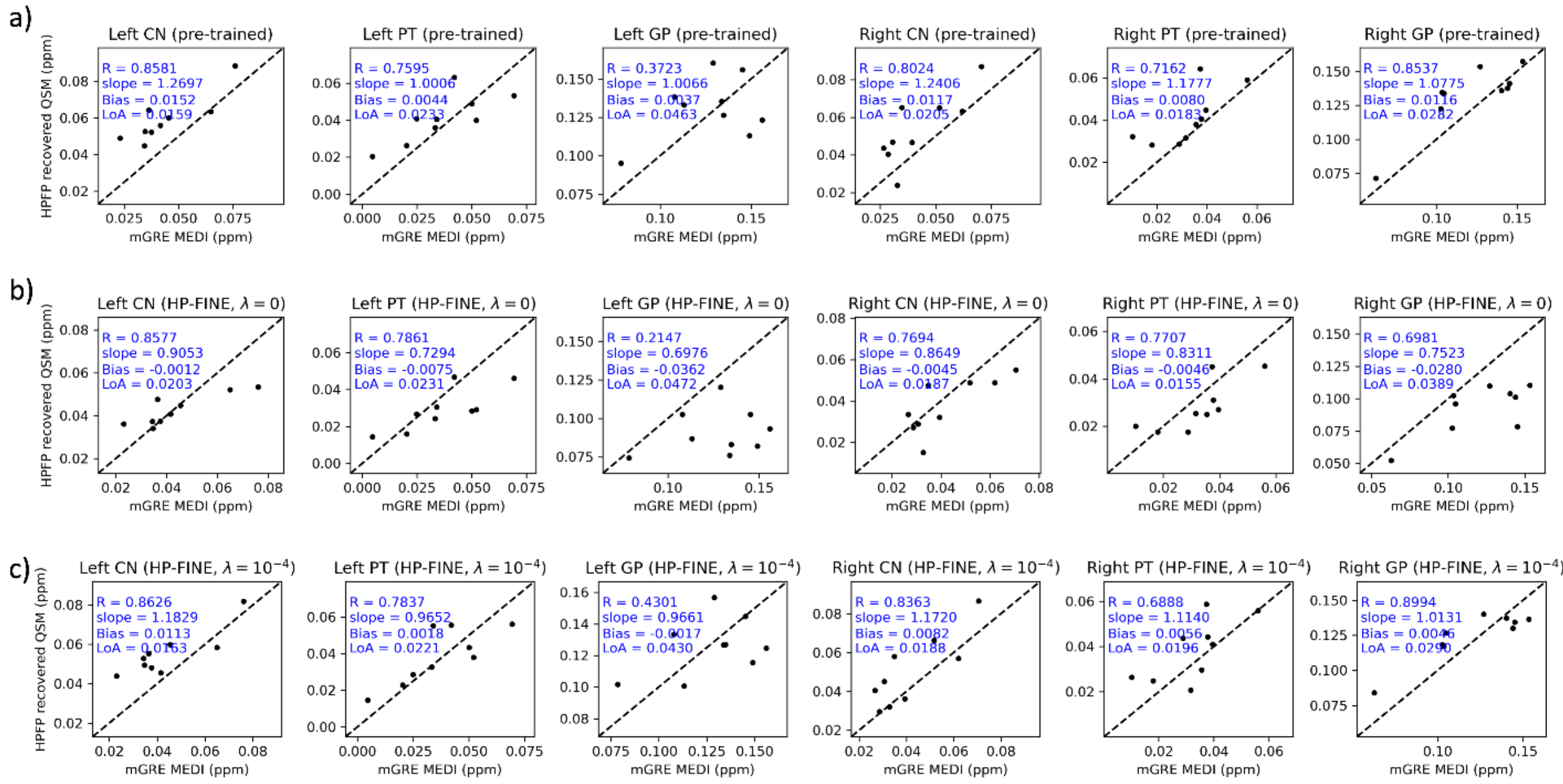

**Figure 5.** Statistics of ROI value linear regression and Bland-Altman analysis conducted on the prospective test dataset using Prognet with the RDF output and filtering augmentation. Compared to the pre-trained results (a), HP-FINE with $\lambda = 0$ (b) suffered from under-estimating ROI values, which were preserved by HP-FINE with $\lambda = 10^{-4}$ (c).

## SUPPORTING INFORMATION

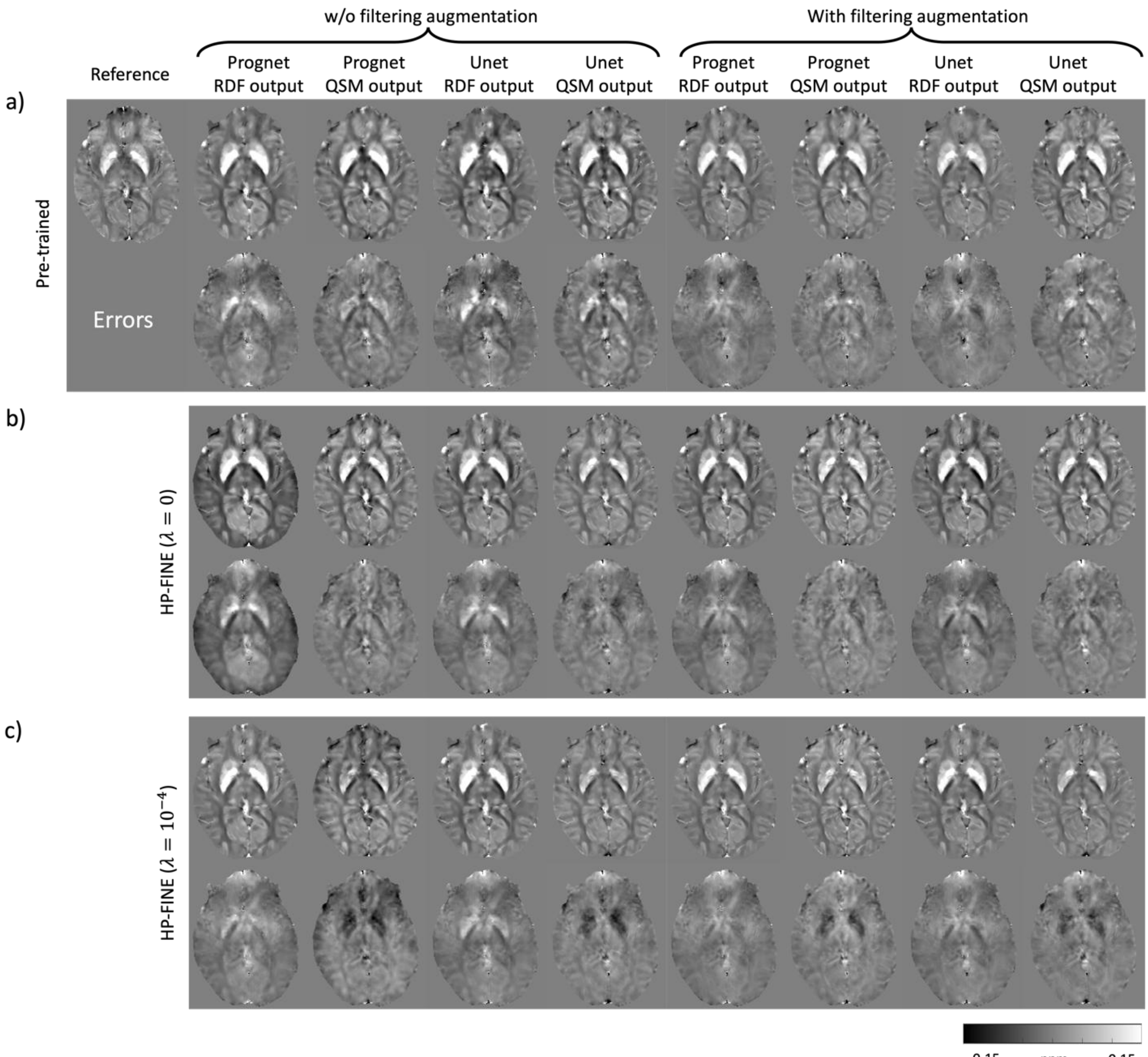

Figure S1. QSM predictions by Unet and Prognet from the HPFP data using the Hann filter with $FC = 1/4$. The comparison between the RDF and QSM outputs, as well as between pre-training with and without the filtering augmentation, was displayed. For the RDF output, both the filtering augmentation and HP-FINE with $\lambda = 10^{-4}$ contributed to improving accuracy. For the QSM output, the filtering augmentation improved accuracy, whereas HP-FINE with $\lambda = 10^{-4}$ resulted in moderate underestimation of the globus pallidus.

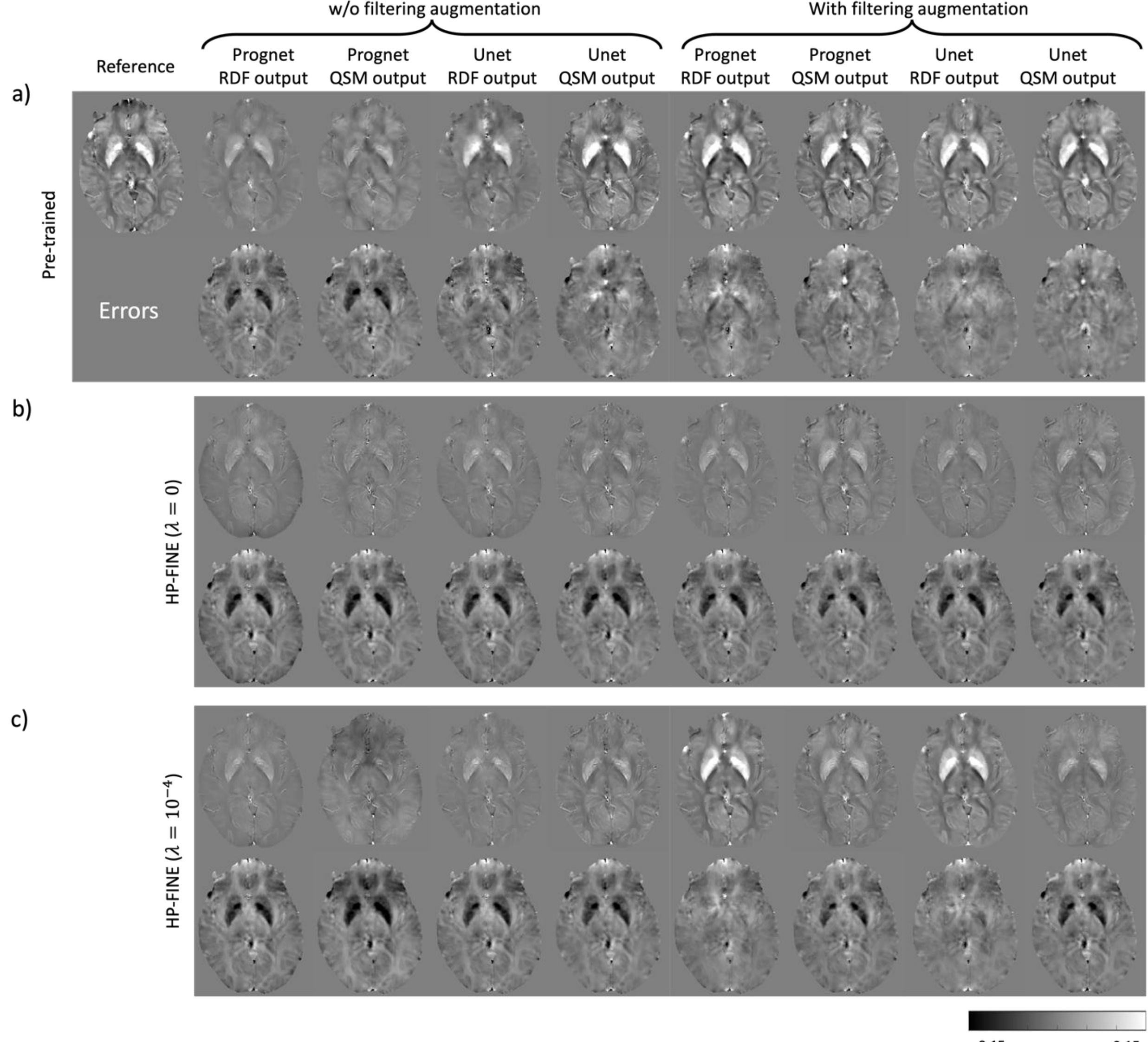

Figure S2. QSM predictions by Unet and Prognet from the HPFP data using the Hann filter with $FC = 3/4$. The comparison between the RDF and QSM outputs, as well as between pre-training with and without the filtering augmentation, was displayed. The filtering augmentation improved the accuracy of the pre-trained results in both the RDF and QSM outputs. For the RDF output, HP-FINE with $\lambda = 10^{-4}$ further improved the accuracy along with the filtering augmentation. For the QSM output, an underestimation of QSM was observed in both HP-FINE with $\lambda = 0$ and $\lambda = 10^{-4}$.

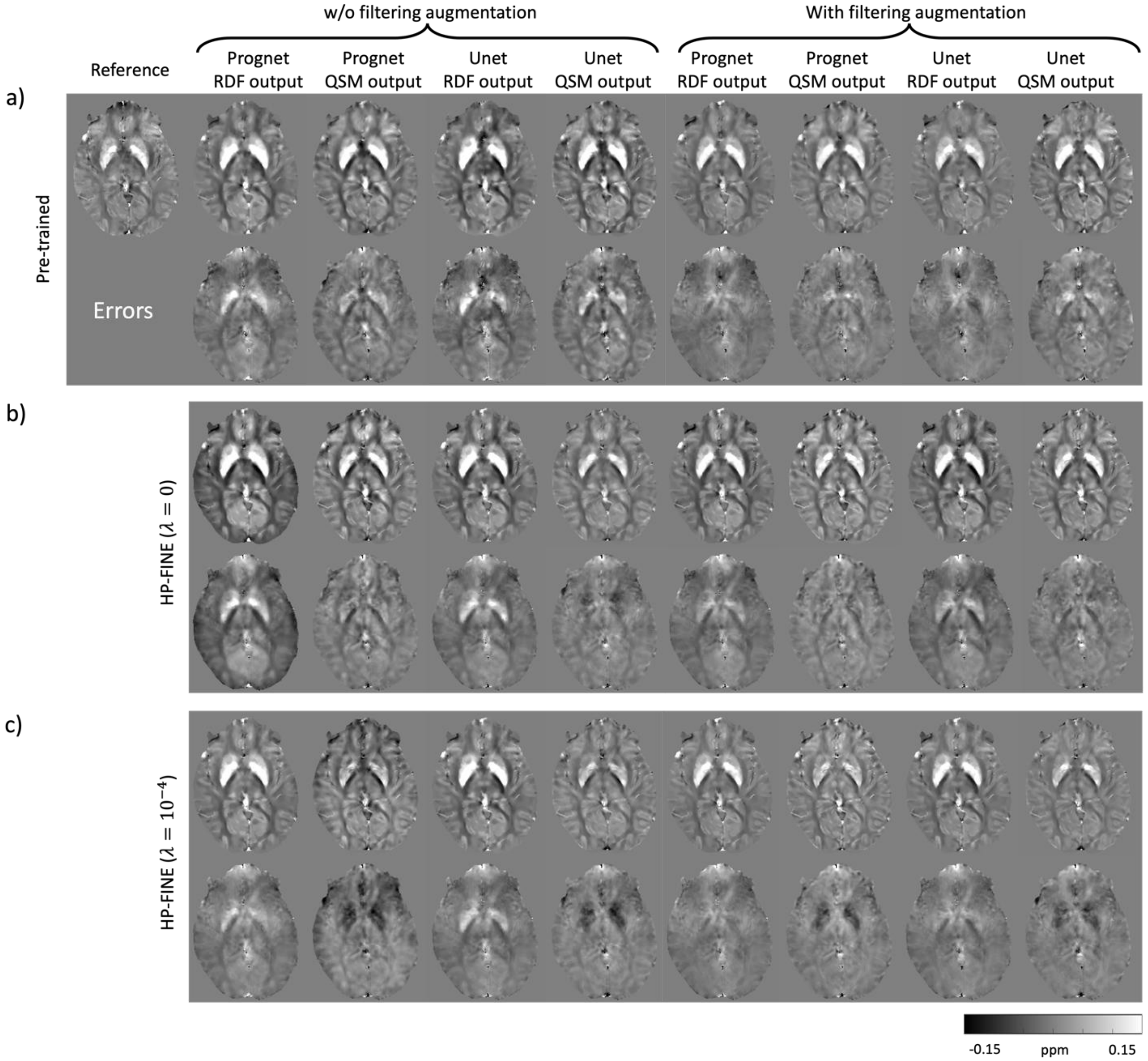

Figure S3. QSM predictions by Unet and Prognet from the HPFP data using the Gaussian filter with $FC = 1/4$. Similar to the results of the Hann filter in Figure S1, both the filtering augmentation and HP-FINE with $\lambda = 10^{-4}$ contributed to improving accuracy for the RDF output, whereas only the filtering augmentation improved accuracy for the QSM output.

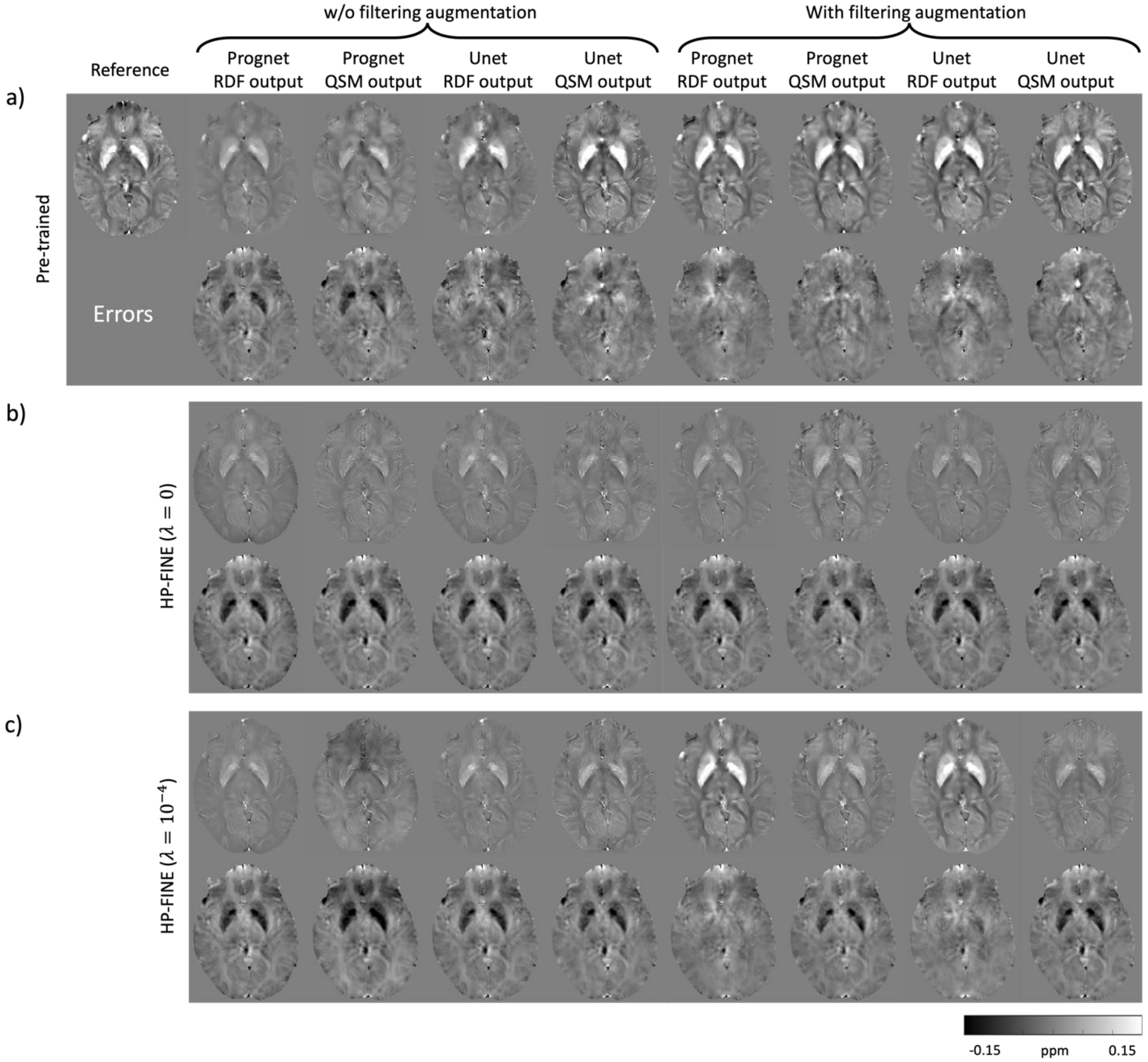

Figure S4. QSM predictions by Unet and Prognet from the HPFP data using the Gaussian filter with $FC = 3/4$. Similar to the results of the Hann filter in Figure S2, the improved accuracy achieved by adding the filtering augmentation was further improved with HP-FINE at $\lambda = 10^{-4}$ for the RDF output. An underestimation of QSM was observed in both HP-FINE with $\lambda = 0$ and $\lambda = 10^{-4}$ for the QSM output.

S5 a):

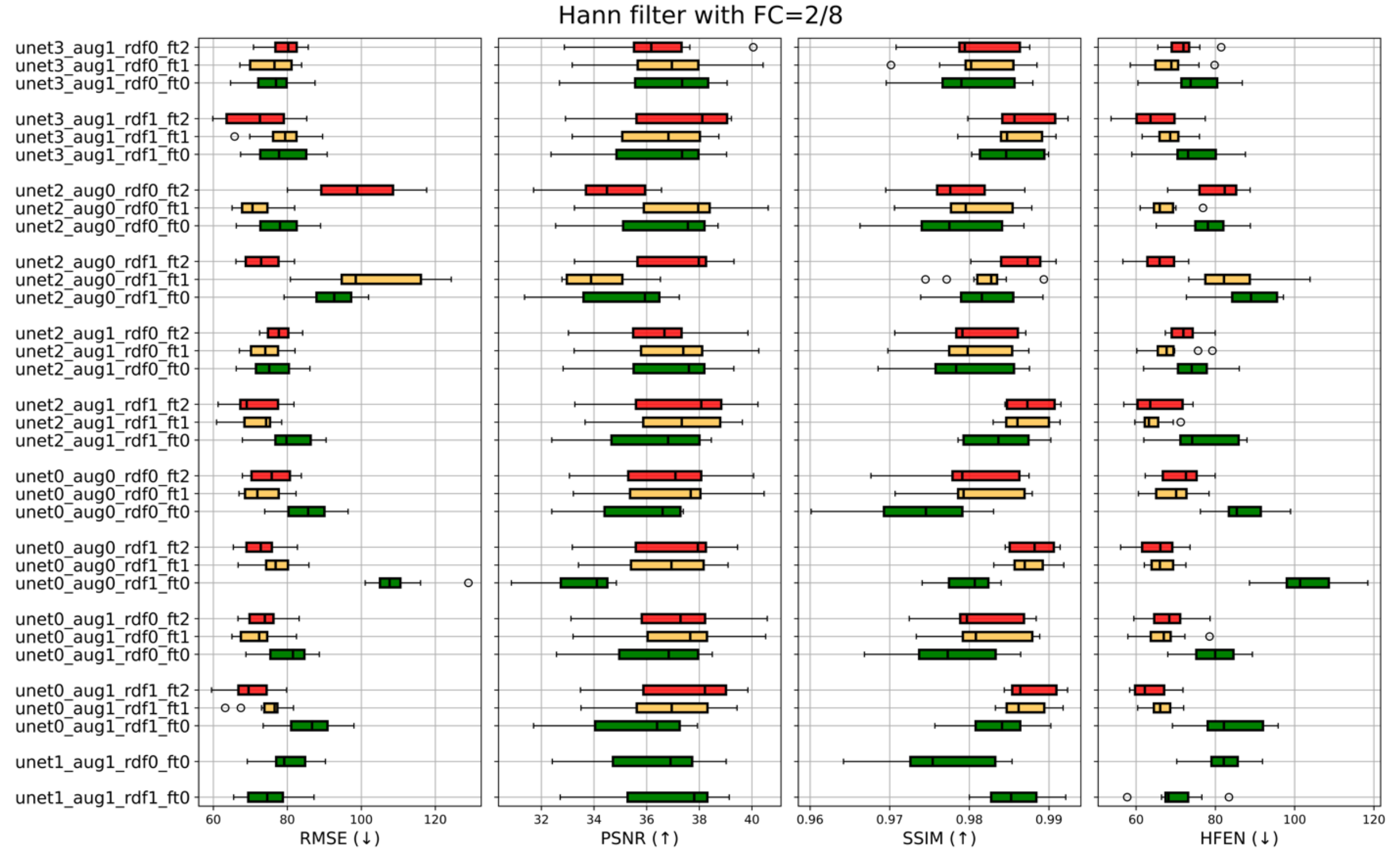

S5 b):

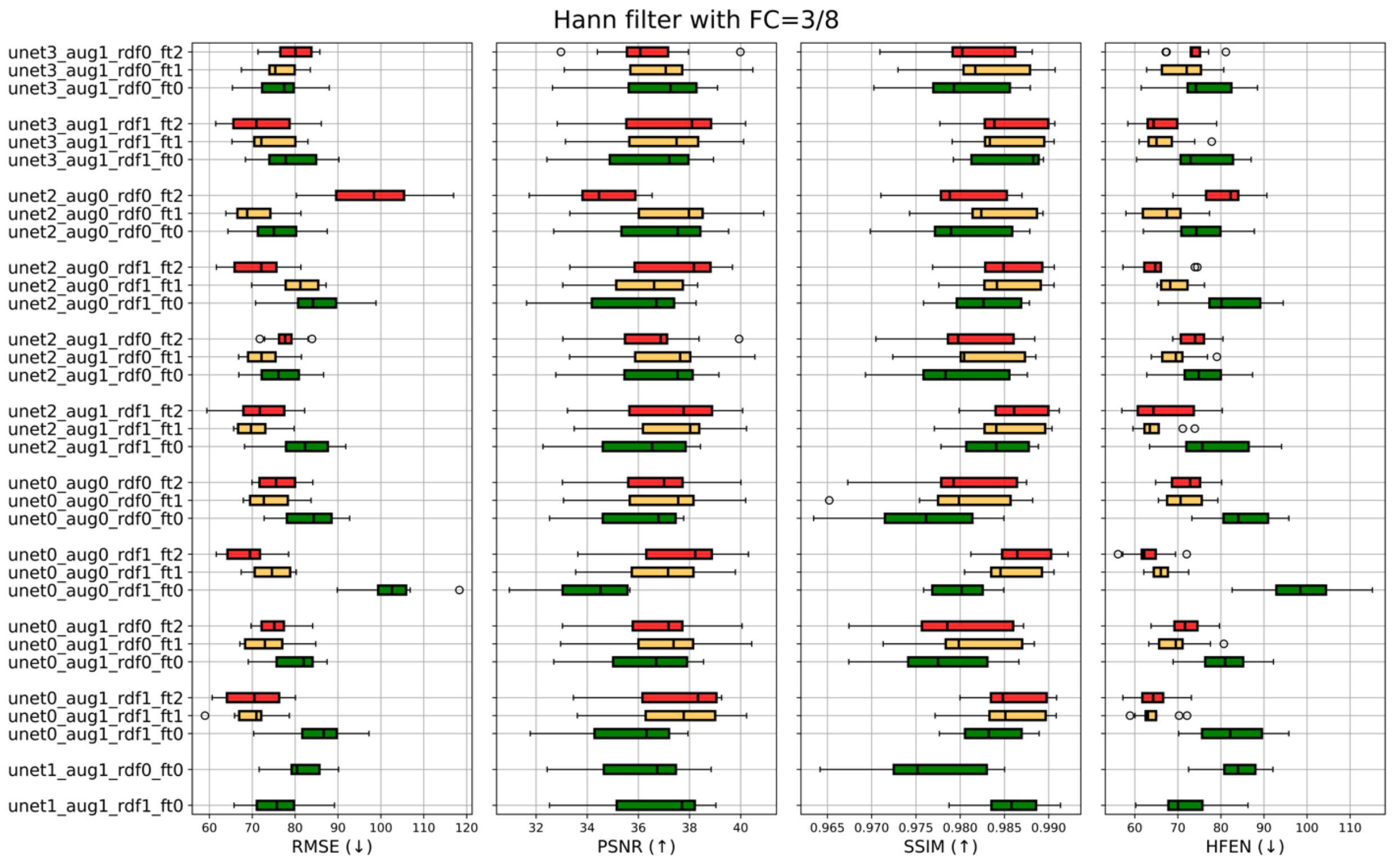

S5 c):

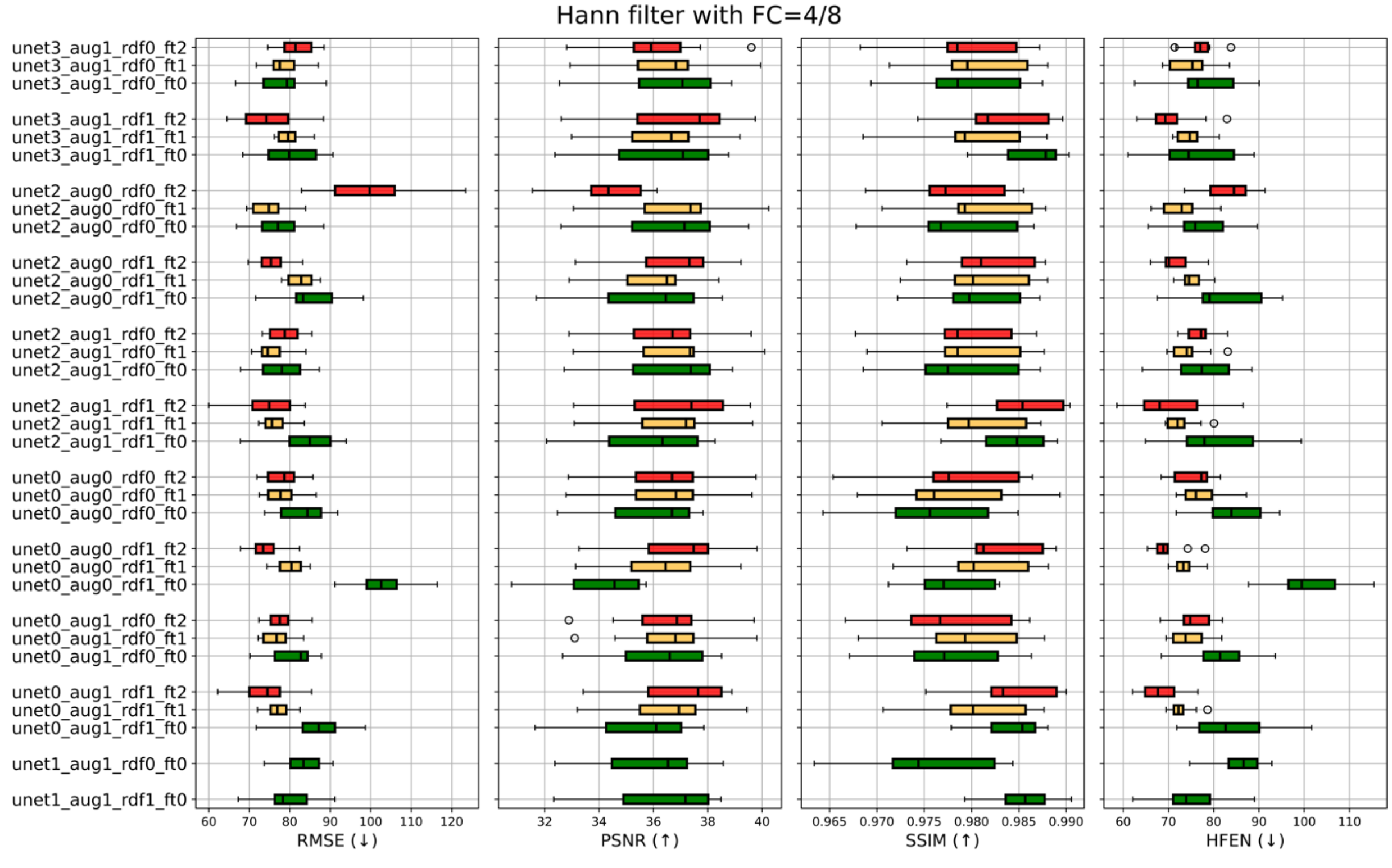

S5 d):

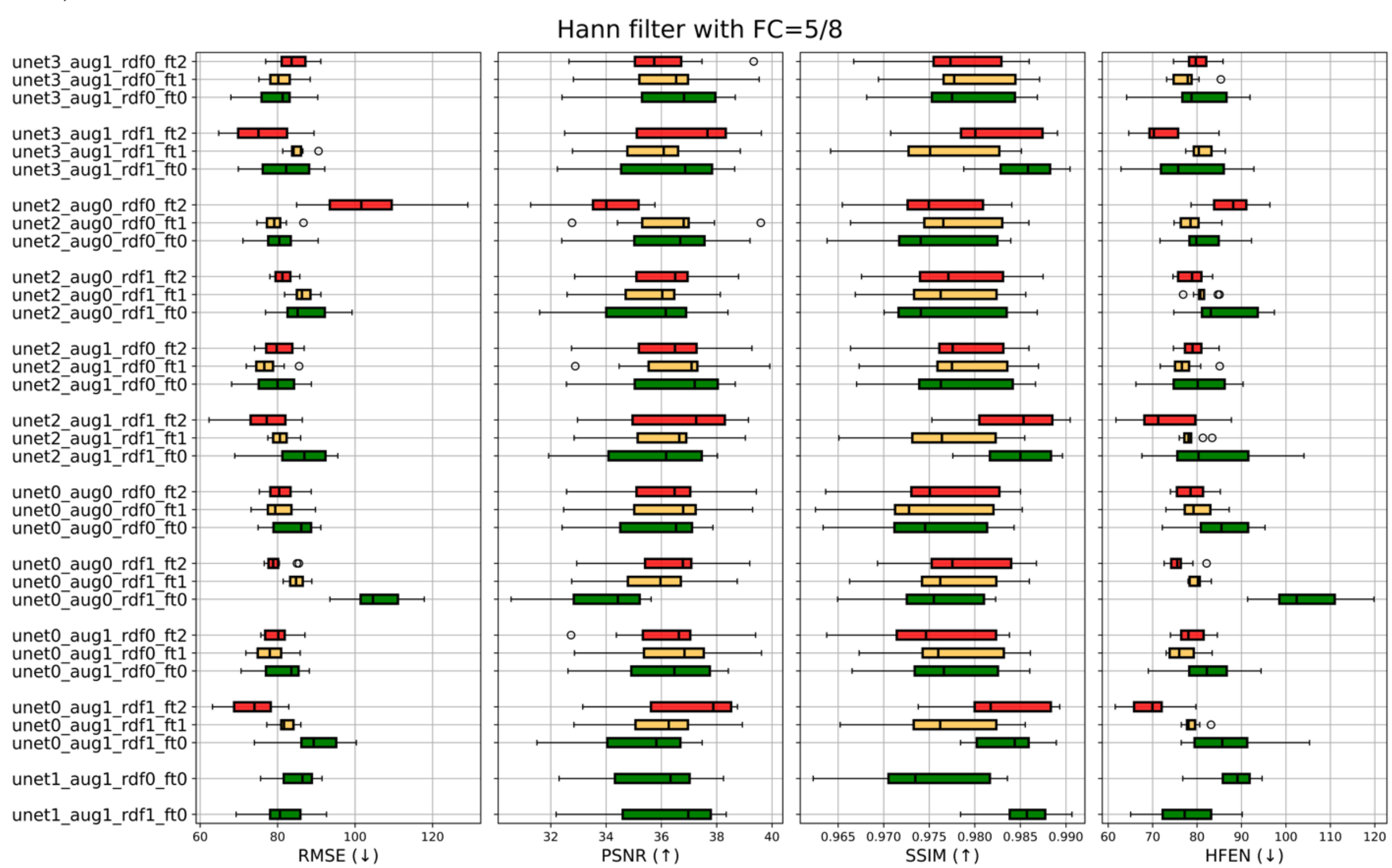

S5 e):

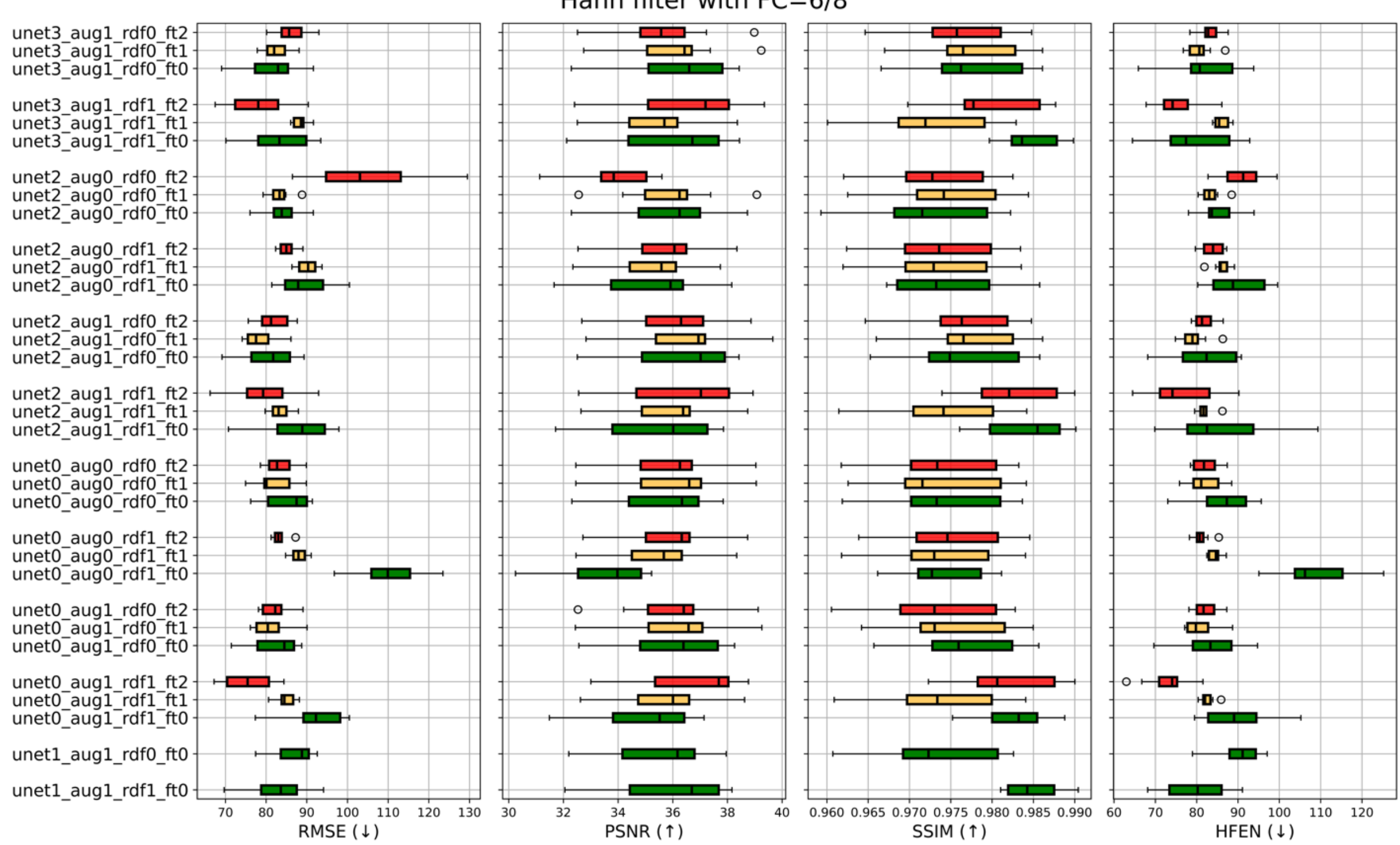

Figure S5. The boxplot of quantitative metrics on the test datasets using the Hann filters with $FCs = 1/4, 3/8, 1/2, 5/8,$ and $3/4$. Overall, the filtering augmentation, RDF output, HP-FINE with $\lambda = 0$, and HP-FINE with $\lambda = 10^{-4}$ all contributed to improved accuracy. Among these, HP-FINE made a more substantial contribution by improving from the pre-trained results, with HP-FINE at $\lambda = 10^{-4}$ slightly outperforming $\lambda = 0$ in most cases for the RDF output. Abbreviations used: unet0: Unet, unet1: Big Unet, unet2: Prognet pre-trained with a sum of losses, unet3: Prognet pre-trained with a single loss, aug0: without the filtering augmentation for pre-training, aug1: with the filtering augmentation for pre-training, rdf0: QSM output, rdf1: RDF output, ft0: pre-trained, ft1: HP-FINE with $\lambda = 0$, and ft2: HP-FINE with $\lambda = 10^{-4}$.

S6 a):

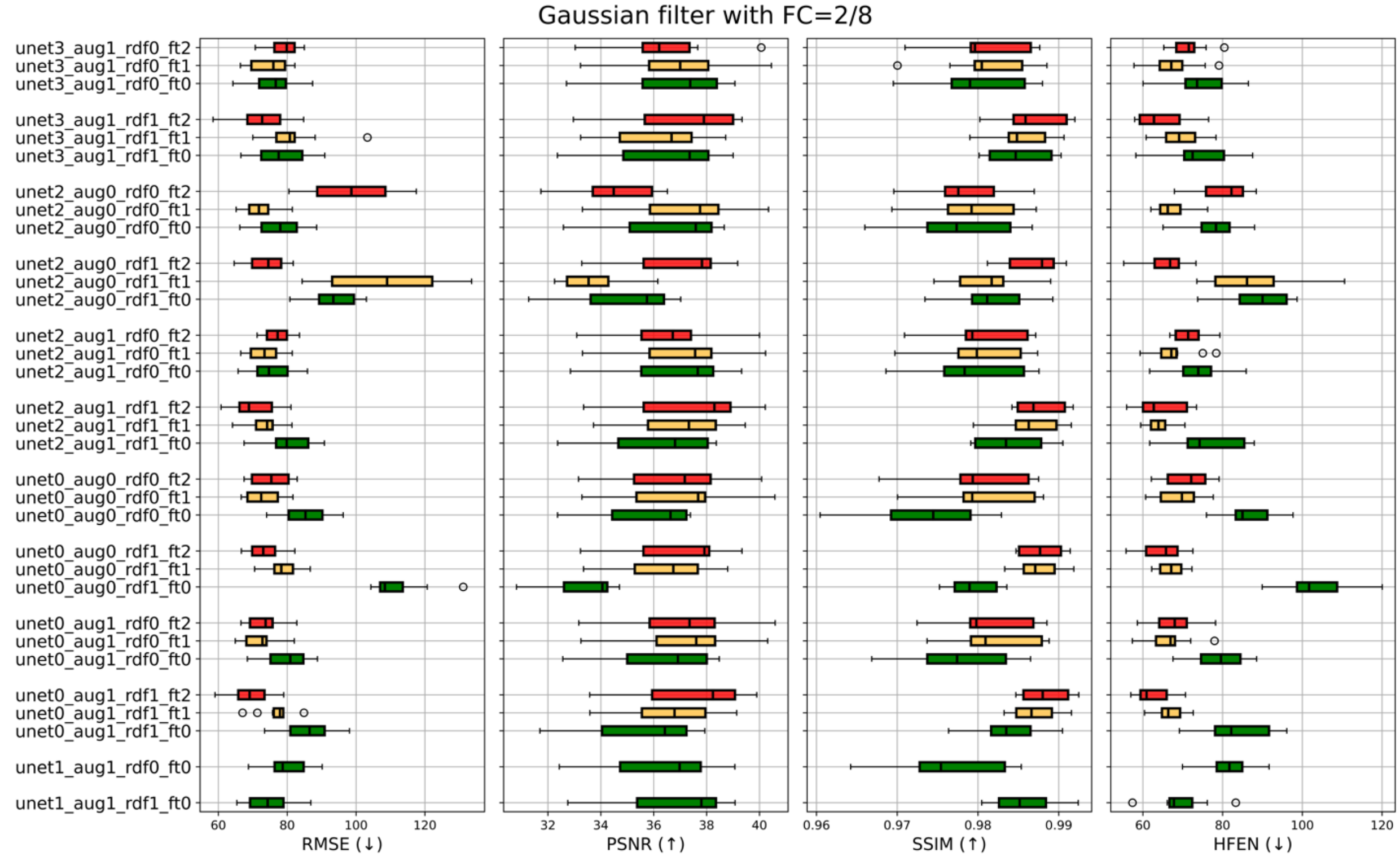

S6 b):

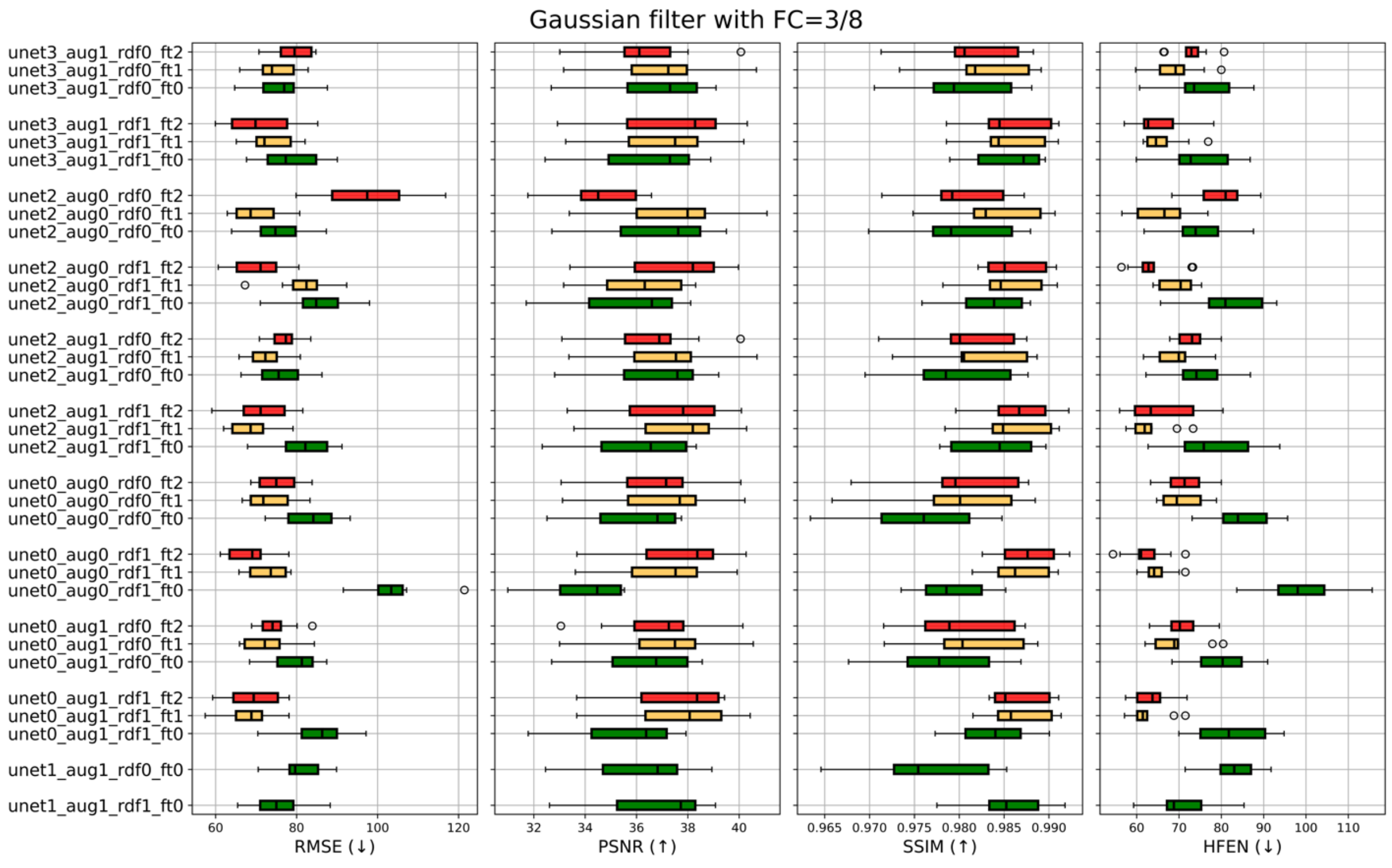

S6 c):

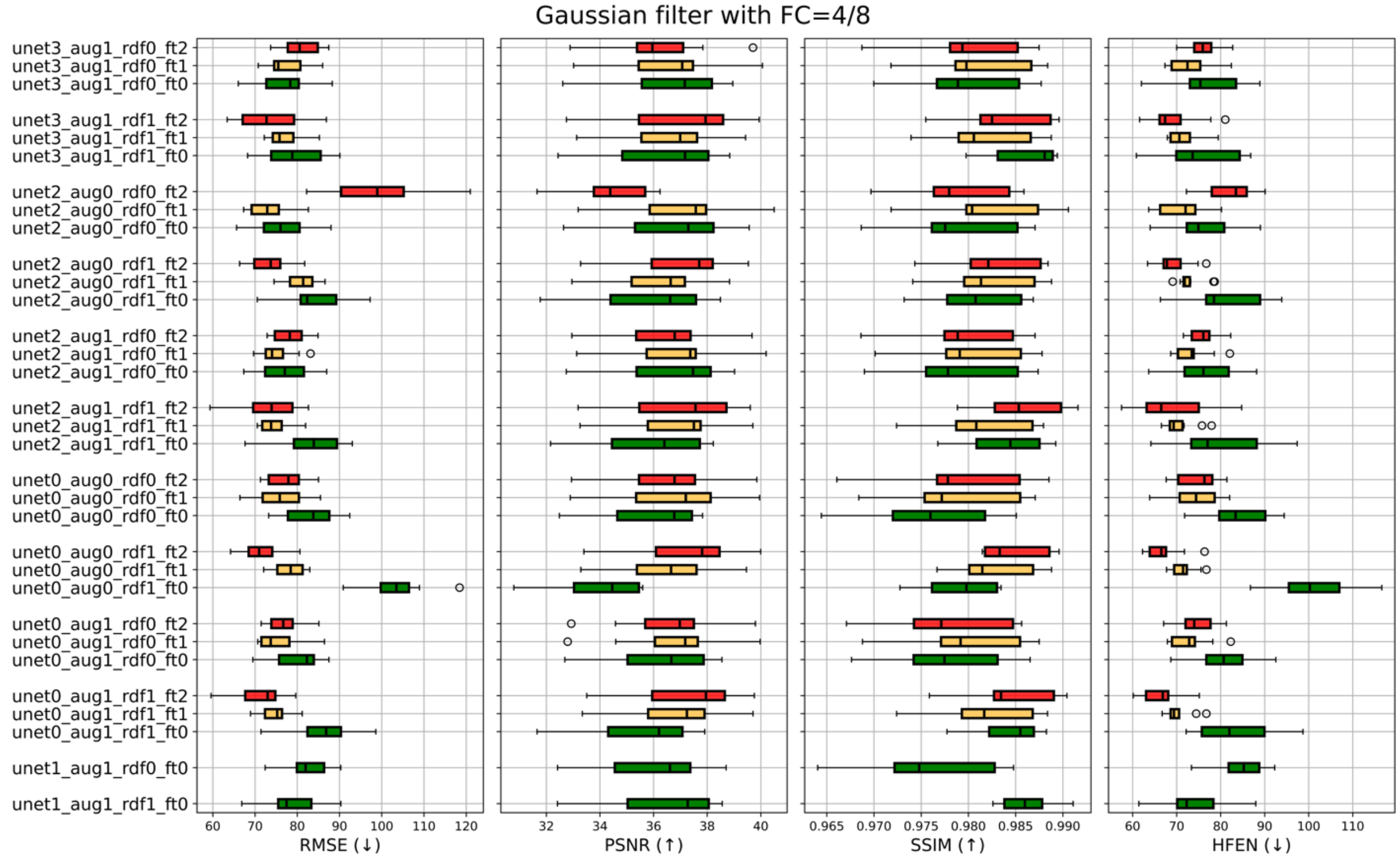

S6 d):

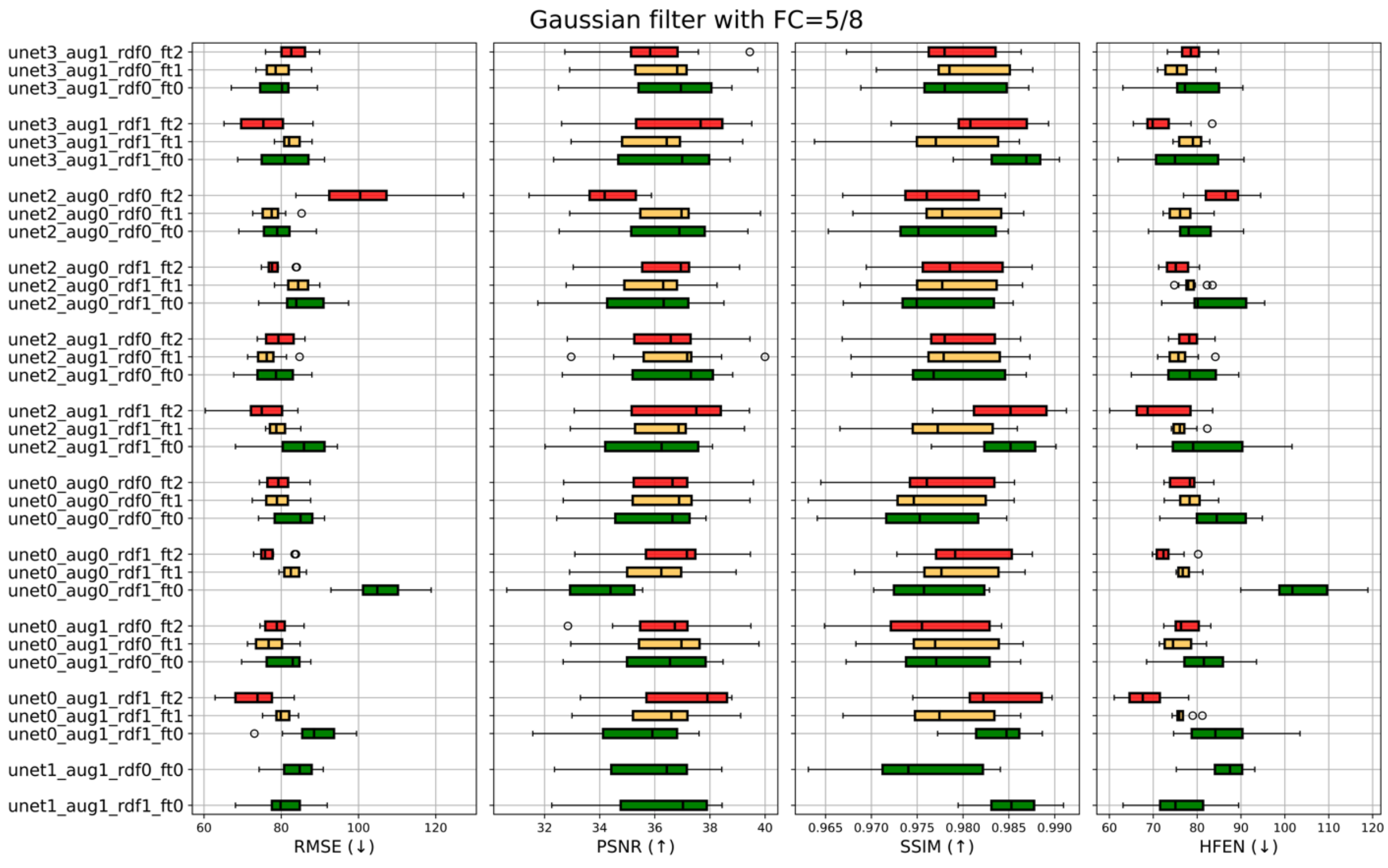

S6 e):

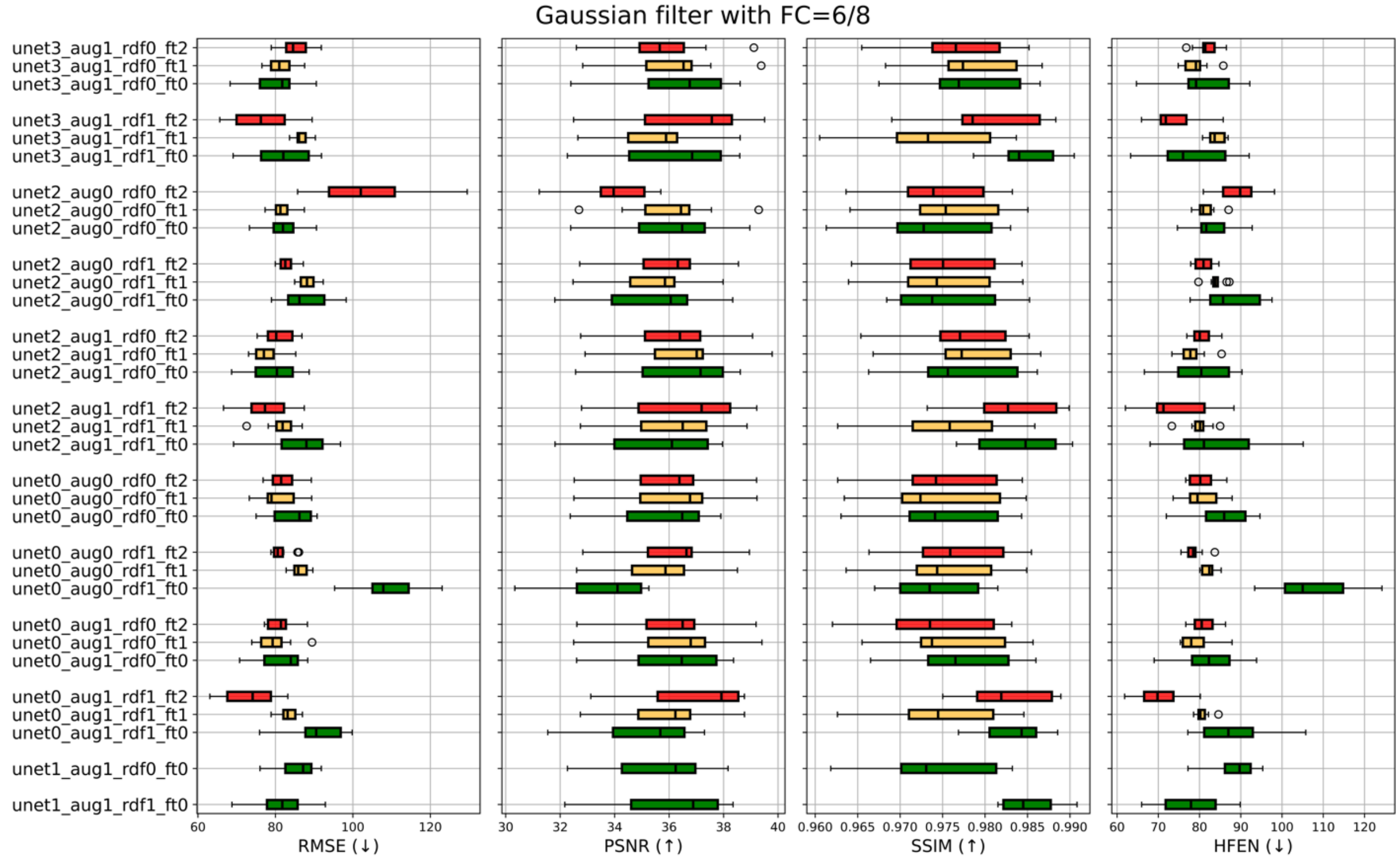

Figure S6. The boxplot of quantitative metrics on the test datasets using the Gaussian filters with $FCs = 1/4, 3/8, 1/2, 5/8,$ and $3/4$. Similar to the results of the Hann filters in Figure S5, the filtering augmentation, RDF output, and HP-FINE with $\lambda = 0$ and $\lambda = 10^{-4}$ all contributed to an overall improvement in accuracy. HP-FINE at $\lambda = 10^{-4}$ slightly outperformed $\lambda = 0$ in most cases for the RDF output. Abbreviations used can be found in the legend of Figure S5.

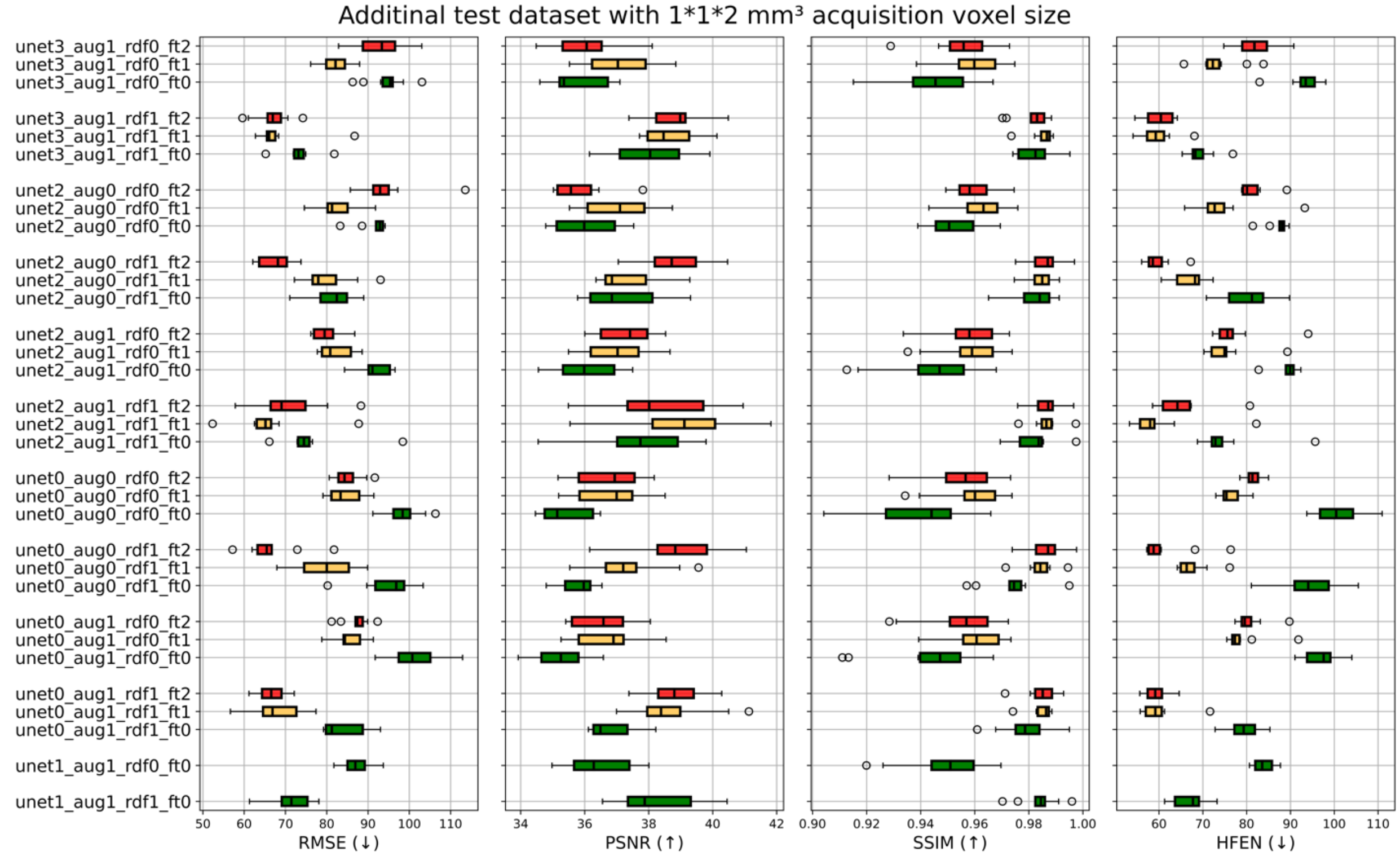

Figure S7. The boxplot of quantitative metrics on the test data acquired with another voxel size. Compared to the RDF output, all methods exhibited a noticeable performance decrease with the QSM output. HP-FINE with $\lambda = 0$ and $\lambda = 10^{-4}$ significantly improved the accuracy of all pre-trained networks. Best performing methods were the ones using the RDF output and HP-FINE with $\lambda = 10^{-4}$. Abbreviations used can be found in the legend of Figure S5.

Table S1. Number of parameters and computational time for various architectures. An increase in computational time averaging 12 seconds was observed on Unet and Prognet during HP-FINE, compared to an average of 1 second for inference without fine-tuning.

| **Architecture Name** | **Number of parameters (millions)** | **Average inference time of pre-trained models (seconds)** | **Average inference time of HP-FINE (seconds)** |
|---|---|---|---|
| Unet | 22.5 | 0.6 | 12.0 |
| Progressive Unet | 90.3 | 1.0 | 12.0 |
| Big Unet | 90.3 | 1.0 | N/A |

**Supplementary spreadsheet.** Statistics of ROI value linear regression and Bland-Altman analysis for all methods on the prospective test dataset. For each network, the RDF output with HP-FINE at $\lambda = 10^{-4}$ outperformed that at $\lambda = 0$ in terms of preserving ROI values, a result not achieved by the QSM output at either $\lambda = 10^{-4}$ or $\lambda = 0$. Prognet outperformed Unet and Prognet (single loss) in the preservation of ROI values. Abbreviations used can be found in the legend of Figure S5.

## APPENDIX A

Low-pass filter $L_{FC}(c)$ is defined as:

$$L_{FC}(c) = F^{-1} W_{FC} F c$$

where $F$ and $F^{-1}$ are Fourier and inverse Fourier transforms, and $W_{FC}$ is a Homodyne low-pass window with Fourier space cutoff frequency $0 \leq FC \leq 1$. For image slice $c$ with size $N \times N$ and a distance $n$ to the center of the $N \times N$ matrix, three types of Homodyne windows were used (16):

**Hann**

$$W_{FC}(n) = \begin{cases} 0.5 + 0.5\cos\left(\frac{2\pi n}{N * FC}\right), & \frac{2n}{N * FC} \leq 1 \\ 0, & otherwise \end{cases}$$

**Hamming**

$$W_{FC}(n) = \begin{cases} 0.54 + 0.46\cos\left(\frac{2\pi n}{N * FC}\right), & \frac{2n}{N * FC} \leq 1 \\ 0, & otherwise \end{cases}$$

**Gaussian ($\sigma = 0.4$)**

$$W_{FC}(n;\sigma) = \begin{cases} \exp\left(-\frac{\left(\frac{2n}{N * FC}\right)^2}{2\sigma^2}\right), & \frac{2n}{N * FC} \leq 1 \\ 0, & otherwise \end{cases}$$

## APPENDIX B

### Unet

The Unet architecture in this work used "Convolution + ReLU activation + Batch normalization" module as the convolutional layer. Two convolutional layers were sequentially connected at each level of the Unet, with the number of channels at levels 0, 1, 2, 3, and 4 (increasing with higher level indices) set as 32, 64, 128, 256, and 512, respectively. Skip concatenation from the decoder to the encoder was applied at each level of the Unet.

### Progressive Unet (Prognet)

The Prognet architecture comprised two Unets. The first Unet took the HPFP as input and outputted the RDF or QSM. The second Unet accepted the concatenated input of HPFP and the outputted RDF or QSM from the preceding Unet to produce the subsequent RDF or QSM output. The second Unet was repeated 3 times to generate the final output.

### Big Unet

The number of convolutional channels at levels 0, 1, 2, 3 and 4 were 64, 128, 256, 512, and 1024, respectively.